\documentstyle [12pt,epsf]{article}

\def\a{\alpha }

\begin{document}

%\begin{titlepage}
\vspace{2cm}

\begin{center}
{\Large\bf QCD analysis of $xF_3$ structure function data and
power correction to $\alpha_s$
%the coupling constant
}
\end{center}
\vskip 16 mm

\begin{center}
{\bf A.V.~Sidorov}\\
{\it Bogoliubov Laboratory of Theoretical Physics\\
 Joint Institute for Nuclear Research, 141980 Dubna, Russia\\
}
\vskip 0.5cm
\end{center}

\vskip 0.5cm
\begin{abstract}
Power corrections both to the strong coupling constant and to the structure
function itself are estimated on the basis of the LO, NLO and
NNLO QCD analysis of $xF_3$ structure function data .
The sign of correction to the coupling constant is found to be negative.
The x-shape of a
higher twist contribution to the structure function is stable up to NNLO.

\vskip 0.5 cm
PACS numbers: 12.38.Bx;~12.38.-t;~13.85.Hd~;14.20.Dh \\
\end{abstract}
\vskip 0.5 cm

%\end{titlepage}

%\newpage

\setcounter{page}{1}

%%{\bf 1. Introduction.}
\vskip 4mm

Power corrections to perturbation QCD predictions both for $Q^2$
-evolution of the running
coupling constant and the structure function itself have
intensively discussed recently \cite{powercorr,AZ9710257}.
At present, precise measurements of structure functions (SF)
and detailed theoretical calculations of QCD predictions for scaling violations
provide an important means of accurate comparison of QCD with experiment.
On this basis,
%The importance of higher--twist (HT) contribution to SF was
%pointed from the very beginning
%of QCD comparison with experimental data \cite{AbbBar} on SF.
during the last years the important information on
the x-shape of power corrections
to the structure function of a nucleon was obtained by the
QCD analysis of data on deep-inelastic scattering of
leptons \cite{VM} - \cite{KKPS2}.

In the present note, detailed LO, NLO and NNLO QCD analyses of precise
$xF_3$ structure function data \cite{CCFR97} have been carried out
in order to
evaluate the power correction to the QCD running coupling
constant

\begin{equation}
\alpha_S(Q^2)=\alpha_S^{pQCD}(Q^2)+\frac{A_2}{Q^2}//
\label{PowCor}
\end{equation}
where in NNLO the coupling constant $\alpha_S^{pQCD}(Q^2)$ can be expressed
in terms
of inverse powers of
  $L=\ln(Q^2/\Lambda_{\overline{MS}}^2)$
as
\begin{eqnarray}
\alpha_S^{pQCD}(Q^2)&=&\frac{1}{\beta_0L}
-\frac{\beta_1 ln(L)}{\beta_0^3 L^2} \\ \nonumber
&&+\frac{1}{\beta_0^5 L^3}[\beta_1^2 ln^2 (L)
-\beta_1^2 ln(L) +\beta_2\beta_0-\beta_1^2]
\end{eqnarray}

Notice that
%\begin{eqnarray}
$\beta_0=11-0.6667f$,
$\beta_1=102-12.6667f$,
$\beta_2=1428.50-279.611f+6.01852f^2$.

We perform the QCD analysis using the method of
the Jacobi polynomial expansion of
 structure functions. This method of solution of the DGLAP equation was
proposed in
\cite{PS} and developed both for unpolarized  \cite{Kretal}
and polarized cases \cite{LSS}. The main formula of this method
allows
approximate reconstruction  of the structure function through a
finite number of Mellin moments of the structure function

\begin{equation}
xF_{3}^{N_{max}}(x,Q^2)=\frac{h(x)}{Q^2}+x^{\alpha}(1-x)^{\beta}
%\times \\ \nonumber
\sum_{n=0}^{N_{max}}
\Theta_n ^{\alpha , \beta}
(x)\sum_{j=0}^{n}c_{j}^{(n)}{(\alpha ,\beta )}
M_{j+2}       \left ( Q^{2}\right )
\label{Jacobi}
\end{equation}

  The $Q^2$-evolution of $ M_N(Q^2)$
is defined by the perturbative QCD

\begin{eqnarray}
M^{QCD}_N(Q^2)
& =& \left [ \frac{\alpha _{S}\left ( Q_{0}^{2}\right )}
{\alpha _{S}\left ( Q^{2}\right )}\right ]^{d_{N}}
H_{N}\left (  Q_{0}^{2},Q^{2}\right )
M_N^{QCD}(Q^2_0) ,~~~N = 2,3, ... \label{m3q2} \\
{}\nonumber \\
d_N & = & \gamma^{(0),N}\bigg/2\beta_0
%\beta_0=(11-\frac{2}{3}f)
        \nonumber
\end{eqnarray}
 Here ~$\a_s(Q^2)$~ is the          strong interaction  constant,
~$\gamma^{(0),N}$~ are nonsinglet leading order anomalous dimensions,
and the factor ~$H_{N}\left (  Q_{0}^{2},Q^{2}\right )$~ contains
 next and next-to-next
to leading order QCD corrections.
Power corrections to the coupling constant are introduced formally
in accordance with (\ref{PowCor}).

\par
Unknown coefficients $M_N^{QCD}(Q^2_0)$ in (\ref{m3q2}) could be parametrized
as the Mellin moments of some function:
\begin{eqnarray}
M_3^{QCD}(N,Q^2_0)&=&\int_{0}^{1}dx{x^{N-2}}ax^b(1-x)^c,
~~~ N = 2,3, ...
\label{Mellf30}
\end{eqnarray}

The shape of the function $h(x)$ as well as parameters
$A_2$, a, b, c, and $\Lambda_{\overline{MS}}$  are found by fitting
the experimental data on the $xF_3(x,Q^2)$ structure function
\cite{CCFR97}.
Detailed description
of the fitting procedure could be found in \cite{KKPS2}. Both terms $h(x)/Q^2$
and $A_2/Q^2$ are considered as pure phenomenological.
For a possible analytic expression see \cite{ShirSol}.
The target mass corrections
are taken into account to the order $o(M^4_{nucl}/Q^4)$.

The results of the fit are presented in Table 1  and Figure 1.

 The values of $A_2$ in Table 1 are opposite in sign to
the lattice results for NLO and NNLO \cite{Burgio}
$A_2^{NLO}=0.22(2)~GeV^2$ and $A_2^{NNLO}=0.21(2)~GeV^2$ obtained
in different
renormalization schemes. Notices also that in \cite{Burgio}
"lattice data" for the coupling constant were analysed,
whereas in the QCD analysis
of structure functions the anomalous dimensions and coefficient functions
were involved too. Even in the leading order, the first factor in the
right-hand side
of (\ref{m3q2}) reproduces the terms with powers differing from $-2$,
 which initially
appears in (\ref{PowCor}).
However, the absolute value of the parameter $A_2$
and a large value of the scale parameter $\Lambda_{\overline{MS}}$
are in qualitative agreement with \cite{Burgio}. On the other hand,
the negative value of $A_2$ is in agreement with predictions of
\cite{ShirSol}.

\vspace{5mm}

\begin{center}
\begin{tabular}{|l|c|c|c|} \hline   \hline
  &$A_2~[GeV^2]$&$\Lambda_{\overline{MS}}~[MeV]$ &$\chi^2$ \\ \hline \hline
\multicolumn{4}{|c|}{h(x) - free}   \\  \hline
 LO    & -0.261 $\pm$ 0.053    &  1140 $\pm$ 110   & 70.5 / 96\\
 NLO   & -0.130 $\pm$ 0.027    &  788 $\pm$ 92  &   71.7 / 96  \\
 NNLO  & -0.123 $\pm$ 0.017   &   561 $\pm$ 46  &   73.4 / 96  \\  \hline  \hline
\multicolumn{4}{|c|}{h(x)=0}   \\  \hline
 LO    & -0.137 $\pm$ 0.011    &   834 $\pm$  31   &115.6 / 96\\
 NLO   & -0.049 $\pm$ 0.012    &  584 $\pm$ 69  &  116.5 / 96  \\
 NNLO  & -0.046 $\pm$ 0.013   &   561 $\pm$ 74  &  103.2 / 96  \\  \hline
 NLO   & -0.011 $\pm$ 0.008    &  267 $\pm$ 36  &  135.4 / 96  \\
 NNLO  & -0.023 $\pm$ 0.005   &   290 $\pm$ 36  &  125.5 / 96  \\  \hline  \hline
\multicolumn{4}{p{9cm}}{{\bf Table 1.}
The results of the LO, NLO
($N_{Max}=10$)
and NNLO  ($N_{Max}=6$)
QCD fit (with TMC) of $xF_3$ data \cite{CCFR97}.
($Q^2_0=3~GeV^2$, $Q^2 > 3~GeV^2$, f=4). The bottom two lines correspond to
substitution of (\ref{PowCor}) into the moments of the
coefficient function only.
}
\end{tabular}
\end{center}

The values of constants $A_2$ for NLO and NNLO are approximately the same
in agreement with the statement that the $1/Q^2$ corrections
to all orders in $\alpha_s$ are of the same order
\footnote{The LO result should not be
considered on this matter
because a nontrivial contribution of the coefficient function to
evolution of the structure function starts from NLO.}  \cite{AZ9710257}.

The shape of h(x) slightly differs
from the results of analysis in Ref. \cite{KKPS2} with $A_2=0$.
The effect of decreasing the power correction to the
structure function \cite{SidHTComm,KKPS2} while going from LO to NNLO
of the perturbative QCD does not exist, as canbeseen from Fig.1.

A special fit for the case $h(x)=0$ gives a negative value for the
parameter $A_2$.
  The increase of the $\chi^2$ parameter
shows that using only power corrections to coupling constant,
one could not reach a good description of experimental data and
it is necessary to introduce additional power corrections
to the structure function itself. A large difference between the value of $A_2$
for $h(x)=0$ and $h(x)\neq 0$ indicates
strong correlations between two power terms:
$A_2/Q^2$ and $h(x)/Q^2$.

Even higher $\chi^2$ is obtained when the formal
substitution of (\ref{PowCor}) is applied
to coefficient finctions and is not applied
to the anomalous-dimention-dependent factor of structure function moment.
This result is presented in the bottom two lines of Table~1.
The values of $A_2$ are small and negative, but parameter the
$\Lambda_{\overline{MS}}$ is in agreement
with \cite{KKPS2}.

In conclusion it should be noted that the presented values of $A_2$
should be taken with a great caution.
One should make use od strict analitic expressions for
the $Q^2$ evolution of the structure function
moments with power corrections to coupling
constant for a reliable fit of data. The nuclear and
threshold effects should be taken into account, as well.

{\bf Acknowledgments}
\vskip 3mm

This research was supported
by the Russian Foundation for Basic Research, Grant No
96-02-18897a.
\\

%%\newpage

\newpage
%\vspace{1cm}
%%%%%%%%%%%%%%%%%%%%%%%%%%%%%%%%%%%%

\vspace*{5.0cm}
%%       \hspace*{.5cm}
\epsfxsize=10cm
\epsfysize=8cm
\centerline{\epsfbox{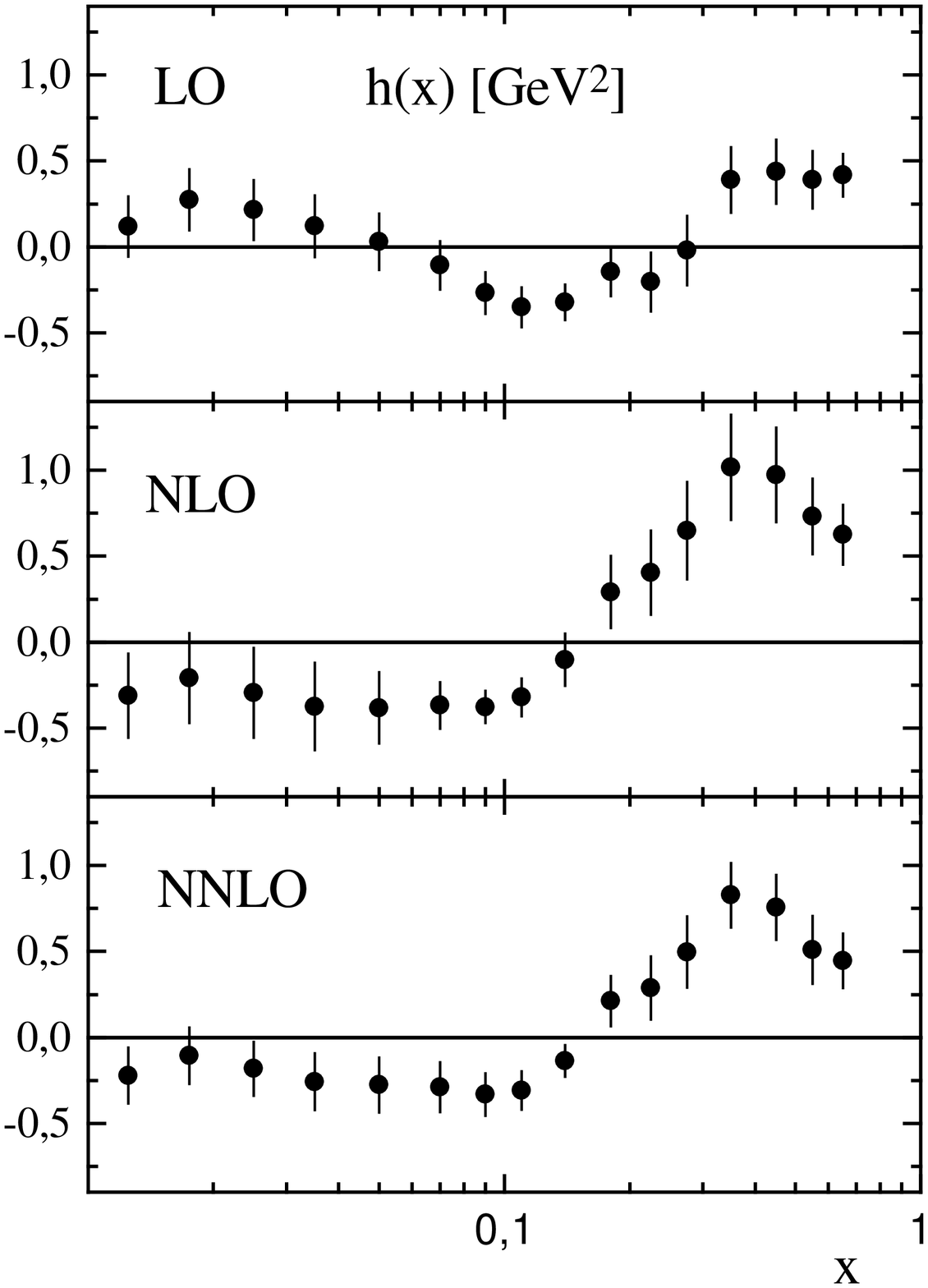}}
%%{\epsfbox{asycompa.eps}}
  \vspace*{-.5cm}
%\begin{center}
Fig.1~The results of the LO, NLO and NNLO
 extractions of twist-4 contributions of $h(x)$.
The power correction to the QCD running coupling
constant $\alpha_S(Q^2)=\alpha_S^{pQCD}(Q^2)+{A_2}/{Q^2}$ is included.
%\end{center}


\begin{thebibliography}{99}

\bibitem{powercorr}
 R. Akhoury and V.I. Zakharov, hep-ph/9705318;\\
 G.Grunberg, hep-ph/9705290,
hep-ph/9705460;\\
M.~Beneke, CERN-TK-98-233, hep-ph/9807443;\\
G. Burgio et al., {\em Phys. Lett.} {\bf B422} (1998) 219.

\bibitem{AZ9710257}
 R. Akhoury and V.I. Zakharov,
{\em Nucl. Phys. Proc. Suppl.} {\bf 64} (1998) 355,
 hep-ph/9710257.
%\bibitem{DGLAP}
%V.N.~Gribov and L.N.~Lipatov, {\em Sov. J. Nucl. Phys.}
%{\bf 15} (1972) 438;
%L.N.~Lipatov, {\em Sov. J. Nucl. Phys.} {\bf 20} (1975) 94;
%G.~Altarelli and G. Parisi, {\em Nucl. Phys.} {\bf B126} (1977) 298;
%see also  Yu.L.~Dokshitzer, {\em JETP} {\bf 46} (1977) 641.

\bibitem{VM}
M.~Virchaux and A.~Milsztajn, {\em Phys. Lett.} {\bf B274} (1992).

%\bibitem{PKK}
%G.~Parente, A.V.~Kotikov and  V.G.~Krivokhizhin,
%{\em Phys. Lett. } {\bf B333} (1994) 190.

\bibitem{SidHT}
 A.V.~Sidorov,
 {\em Phys. Lett.} {\bf B389} (1996) 379.

\bibitem{SidHTComm}
 A.V.~Sidorov,
 JINR Rapid Comm. {\bf 80}
(1996) 11 (hep-ph/9609345).

\bibitem{KKPS2}
A.L. Kataev  et al., {\em Phys. Lett.} {\bf B417} (1998) 374;
[hep-ph/9809500].

\bibitem{CCFR97}
CCFR-NuTeV Collab., W.G. Seligman et al.,
{\em Phys. Rev. Lett.} {\bf 79} (1997) 1213.



\bibitem{PS}
G.~Parisi and  N.~Sourlas, {\em Nucl. Phys.} {\bf B151} (1979) 421.


\bibitem{Kretal}
I.S.~Barker, C.B.~Langensiepen and  G.~Shaw,  {\em Nucl.  Phys.}
{\bf B186} (1981) 61;\\
V.G.~Krivokhizhin et al., {\em Z. Phys.} {\bf C36} (1987) 51,
{\em Z. Phys.} {\bf C48} (1990) 347; \\
A.V.~ Kotikov, G.~ Parente and J.~ Sanchez-Guillen, {\em Z. Phys.}
{\bf C58} (1993) 465;\\
A.L.~Kataev and A.V. Sidorov, {\em Phys. Lett.} {\bf B331} (1994) 179;\\
A.L. Kataev et al., {\em Phys. Lett.} {\bf B388} (1996) 179.

\bibitem{LSS}
E. Leader, A.V. Sidorov and D.B. Stamenov,
hep-ph/9708335 (to be published in IJMPA);
hep-ph/9807251 (to be published in Phys. Rev. D);\\
C. Bourrely et al., Preprint CPT-97/P 3578, hep-ph/9803229.


\bibitem{ShirSol}
 D.V. Shirkov and I.L. Solovtsev,
{\em Phys. Rev. Lett.} {\bf 79} (1997) 1209; \\
D.V. Shirkov, {\em Nucl. Phys. Proc. Suppl.} {\bf 64} (1998) 106,
hep-ph/9708480.

\bibitem{Burgio}
 G. Burgio et al., hep-ph/9809450.



\end{thebibliography}
\end{document}